\documentclass[acmsmall]{acmart}

\usepackage{acronym} %for ac command
\usepackage{algorithm} 
\usepackage{algorithmic}
\usepackage{booktabs}
\usepackage{clrscode3e}
\usepackage{hyperref}
\usepackage{multirow}
\usepackage{graphicx}
\usepackage{subfigure}
\usepackage{textcomp}
\usepackage{xcolor}
\usepackage{tabularx}

\acrodef{HyperGroup}{Hyperedge embedding-based Group recommender}
\acrodef{OGR}{Occasional Group Recommendation}
\acrodef{OG}{Occasional Group}
\acrodef{OGs}{Occasional Groups}
\acrodef{IPM}{Individual Preference Modeling}
\acrodef{THE}{Translating Hyperedge Embedding}
\acrodef{MLP}{Multi-Layer Perception}
\acrodef{HE}{Hyperedge Embedding}
\acrodef{HR}{Hit Ratio}
\acrodef{NDCG}{Normalized Discounted Cumulative Gain}
\acrodef{IPM}{Individual Preference Modeling}
\acrodef{HRL}{Hyperedge Representation Learning}
\acrodef{GNN}{Graph Neural Network}
\acrodef{GR}{Group Recommendation}
\acrodef{DMP}{Decision Making Process}
\acrodef{PGs}{Persistent Groups}
\acrodef{PG}{Persistent Group}
\acrodef{PGR}{Persistent Group Recommendation}

\AtBeginDocument{%
  \providecommand\BibTeX{{%
    \normalfont B\kern-0.5em{\scshape i\kern-0.25em b}\kern-0.8em\TeX}}}

\setcopyright{acmcopyright}
\copyrightyear{2021}
\acmYear{xxx}
\acmDOI{10.1145/0000000.000000}

\acmJournal{TOIS}
\acmVolume{0}
\acmNumber{0}
\acmArticle{0}
\acmMonth{0}

\begin{document}

\title{Hierarchical Hyperedge Embedding-based Representation Learning for Group Recommendation}


\author{Lei Guo}
\email{leiguo.cs@gmail.com}
\affiliation{
  \institution{Shandong Normal University}
%   \streetaddress{No. 1, University Road, Changqing District}
  \city{Jinan}
  \state{Shandong}
  \country{China}
  \postcode{250358}
}

\author{Hongzhi Yin}
\authornote{Corresponding Author.}
\affiliation{%
  \institution{The University of Queensland}
  \city{Brisbane}
  \country{Australia}}
\email{h.yin1@uq.edu.au}

\author{Tong Chen}
\affiliation{%
  \institution{The University of Queensland}
  \city{Brisbane}
  \country{Australia}}
\email{tong.chen@uq.edu.au}

\author{Xiangliang Zhang}
\affiliation{%
  \institution{King Abdullah University of Science and Technology}
  \city{Thuwal}
  \country{Saudi Arabia}}
\email{xiangliang.zhang@kaust.edu.sa}

\author{Kai Zheng}
\affiliation{%
  \institution{University of Electronic Science and Technology of China}
  \city{Chengdu}
  \country{China}}
\email{zhengkai@uestc.edu.cn}

\renewcommand{\shortauthors}{Guo and Yin, et al.}

\begin{abstract}
\ac{GR} aims to recommend items to a group of users. In this work, we study \ac{GR} in a particular scenario, namely \ac{OGR}, where groups are formed ad-hoc and users may just constitute a group for the first time, that is, the historical
group-item interaction records are highly limited.
Most state-of-the-art works have addressed the challenge by aggregating group members' personal preferences to learn the group representation.  However, the representation learning for a group is most complex beyond the aggregation or fusion of group member representation, as the personal preferences and group preferences may be in different spaces and even orthogonal. In addition, the learned user representation is not accurate due to the sparsity of users' interaction data.
Moreover, the group similarity in terms of common group members has been overlooked, which however has the great potential to improve the group representation learning.
In this work, we focus on addressing the aforementioned challenges in group representation learning task, and devise a hierarchical hyperedge embedding-based group recommender, namely HyperGroup. 
Specifically, we propose to leverage the user-user interactions to alleviate the sparsity issue of user-item interactions, and design a \ac{GNN}-based representation learning network to enhance the learning of individuals' preferences from their friends' preferences, which provides a solid foundation for learning groups' preferences.
To exploit the group similarity (i.e., overlapping relationships among groups) to learn a more accurate group representation from highly limited group-item interactions, we connect all groups as a network of overlapping sets (a.k.a. hypergraph), and treat the task of group preference learning as embedding hyperedges (i.e., user sets/groups) in a hypergraph, where an inductive hyperedge embedding method is proposed.
To further enhance the group-level preference modeling, we develop a joint training strategy to learn both user-item and group-item interactions in the same process.
We conduct extensive experiments on two real-world datasets and the experimental results demonstrate the superiority of our proposed HyperGroup in comparison to the state-of-the-art baselines.
\end{abstract}




\maketitle

\section{Introduction}
With the recent advances in social networking services like Meetup and Facebook Event~\cite{liu2019discrete,sun2018attentive}, it is increasingly convenient for people with similar backgrounds (e.g., occupations, hobbies, locations, etc.) to form social groups and participate in activities in groups~\cite{yin2018joint,gao2016collaborative}, such as group tours, class reunion, and family dinners. It is becoming essential to develop group recommender systems~\cite{huang_an,cao_social2019,xiao_fairness_2017,Xiao_disparity_2017} to provide groups with appropriate recommendations (e.g., recommend a restaurant or a concert).

Generally, groups can be divided into persistent groups and occasional groups~\cite{yin2019social,guo2020group,cao2018attentive} based on whether they have stable group members. Concretely, \ac{PGs}~\cite{vinh2019interact,hu2014deep,said2011group} (a.k.a. static groups) often have fixed group members and abundant group-item interactions. For this kind of groups, we can directly apply the recommendation methods designed for individual users~\cite{wang_neural_2019,he2020lightgcn,chen2019air,Guo_streaming_2019} by treating each group as a pseudo user, since there are sufficient persistent group-item interactions. \ac{OGs}~\cite{liu2012exploring,baltrunas2010group,yuan2014generative,guo2020group} (a.k.a. cold-start group) refer to groups that are casually formed by ad-hoc users. 
As such kinds of groups are commonly established for temporary events (such as ride-sharing or attending an academic conference), the historical group-item interaction records are highly limited and even unavailable.
Thus, the representation learning for an \ac{OG} is more challenging than that for a \ac{PG}. 
Moreover, as different group members have different social influences and contribute differently to the group decision, \ac{OGR} is much more complicated than making recommendations to individual users, and static and predefined aggregation methods~\cite{baltrunas2010group,amer2009group,salehi2015preference,berkovsky2010group} are incompetent for the high complexity of group decision-making. In this study, we focus on \ac{OGR}, which is more challenging and also more general in real-world applications compared with Persistent Group Recommendation (PGR).

To address the above challenges,
some advanced data-driven aggregation methods~\cite{cao2018attentive,yin2019social,guo2020group} have been proposed to learn the group representation.
For example, Cao et al.~\cite{cao2018attentive} incorporated the attention mechanism to dynamically aggregate different group members’ preference information. To investigate the impact of group members' social influence on the group decision, Yin et al.~\cite{yin2019social} proposed a social influence-based group recommender to improve the preference aggregation process. But they ignore that the group's final decision is usually reached through consensus among group members, and the social interactions are unexplored. To address this limitation, Guo et al.~\cite{guo2020group} treated the group representation learning process as a multiple voting step, and developed a stacked social self-attention network~\cite{vaswani2017attention} to aggregate group members' preferences. To alleviate the sparsity issue of group-item interaction data, the users' individual activity data has been leveraged to complement the group recommendation task in their method.

However, as the personal preferences and group preferences may be in different spaces and even orthogonal, the representation learning for a group is most complex beyond the aggregation or fusion of group member representation. Moreover, existing works tend to isolate groups when developing preference aggregation strategies, and ignore the fact that driven by multifaceted interests, a user may belong to multiple groups meanwhile. That is, the similarity between groups in terms of common group members is overlooked, which is substantially helpful to enhance the group preference modeling.
Take a toy case for example, suppose we have groups A = \{Amy, Bob, Carl\} and B = \{Amy, Bob, Eric\}. When making recommendations for group A, if we know Amy and Bob are also in B, we can leverage the preference information of group B to improve the inference of group A's preference. Intuitively, the more common members A and B share, the more similar their group preferences will be, which provides another way to alleviate the data sparsity of group-item interaction data. Unfortunately, existing group recommendation methods fail to capture this important group-level similarity.
In addition, these existing works have integrated the individual activity data to alleviate the sparsity issue of group-item interaction data, but they ignore a widely recognized fact that most of the user-item interaction data (i.e., individual activity data) are also extremely sparse.

\begin{figure}
    \centering
    \includegraphics[width=5cm]{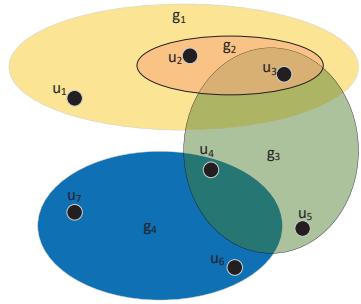}
    \caption{Example of a hypergraph, where $g_i$ denotes the $i$-th group/hyperedge that connects all the users within it. For example, users $u_7, u_6$ and $u_4$ are all connected by group/hyperedge $g_4$. The edge (a.k.a overlapping relationship) between two hyperedges/groups is established if they share at least one common group member, such as hyperedges/groups $g_3$ and $g_4$ share the same user $u_4$, then there is a connection between $g_3$ and $g_4$.   }
    \label{fig:hypergraph_example}
\end{figure}

To this end, instead of simply aggregating personal preferences as the group preference, we propose a hierarchical \acf{HyperGroup} to learn group representation, which consists of \ac{IPM}, \ac{HRL} and joint training components. 
More specifically, to address the sparsity issue of the user-item interactions, in our \ac{IPM} component, we exploit the user-user interactions with the assumption that a user’s preference can be indicated by his/her friends due to the social principle of homophily. Technically, we develop a \ac{GNN}-based node embedding approach to learn group members' personal preferences that provide solid foundations for learning the preferences of \ac{OGs}.
To exploit the group similarity (i.e., overlapping relationships among groups) to enhance the process of learning groups' preferences, in our \ac{HRL} component, we connect all groups as a network of overlapping sets (i.e., hypergraph, an example is shown in Fig.~\ref{fig:hypergraph_example}), and treat group representation learning as embedding hyperedges in a hypergraph, where a hyperedge is an edge that connects all users who belong to the same group. Then,
an innovative inductive hyperedge embedding component is proposed to learn the representation of each hyperedge (i.e., group) by aggregating the representations of its incident hyperedges (i.e., groups that share at least one common member).
Finally, a joint training strategy is developed to optimize the user-item and group-item recommendation tasks simultaneously, which provides an efficient way to accelerate and enhance the group-level preference modeling and learning.

Note that, the method that views a group as a hyperedge is different from the methods that directly treat groups as pseudo users, since they only depend on the highly limited group-item interaction records, while our method additionally considers both user-user and user-item interactions.
Our method is also different from existing \ac{OGR} studies that tend to model each group individually. \ac{HyperGroup} exploits the group similarity in terms of common group members, thus making full use of the group-level collaborative preference signals to enhance the group preference learning. 

The main contributions of this work are listed as follows:
\begin{itemize}
    \item
    In this work, we 1) exploit the group similarity to improve the group representation learning and 2) introduce hyperedges to model groups.
    \item We propose \ac{HyperGroup}, a novel GNN-based model for \ac{OGR} tasks. Specifically, \ac{HyperGroup} addresses the fundamental data sparsity problem in user-item interactions by leveraging friends' preferences from the social network, and further exploits group-level similarity with an innovative hyperedge embedding scheme for learning expressive group representations.
    \item We conduct extensive experiments on two real-world datasets and the experimental results demonstrate the superiority of our proposed method.
\end{itemize}

\section{Related Work}
The group recommendation techniques that have been widely studied in the last years can be divided into two categories~\cite{OConnor2001,quintarelli2016recommending}: recommendations to persistent groups and occasional groups. In a persistent group, it is assumed to have sufficient group feedback, while in an occasional group, the group feedback is highly limited and even not available. The group preference in the latter case must be learned on the basis of those of its members.

\subsection{Persistent Group Recommendation}
Due to persistent groups often have stable members and rich historical interactions~\cite{quintarelli2016recommending,yin2019social}, previous studies~\cite{chen2008_a,hu2014deep,seko_group_2011} on this kind of group were mainly focused on treating groups as pseudo-users, and then adopt conventional personalized recommendation techniques~\cite{chen2019improving,TOIS2019_cite_leizhu} for making group recommendations. For example, to estimate the rating that a group of members might give to an item, Chen et al.~\cite{chen2008_a} proposed a genetic algorithm-based recommendation method by predicting the possible interactions among group members. But this method relies on known group preferences and is only implementable for persistent groups.
Seko et al.~\cite{seko_group_2011} leveraged the entities that characterize groups (e.g., the tendency of content selection and the relationships among group members) to achieve high group recommendation accuracy. However, their method is also only applicable to predefined groups, such as couples and families, and requires a large amount of group behavior history.
To relieve the vulnerability of data~\cite{Bengio2013_representation}, Hu et al.~\cite{hu2014deep} devised a deep architecture model via using high-level features that are learned from lower-level features to represent group preference.
Similar to work~\cite{seko_group_2011}, this method only focuses on pre-defined groups instead of occasional groups, and thus cannot be applied to our setting.

\subsection{Occasional Group Recommendation}
As existing works developed for occasional group recommendation mainly focus on investigating the strategies of aggregating individual preferences to conduct group recommendations, in the following we review these studies from two aspects: late aggregation methods and early aggregation methods.

\subsubsection{Late Aggregation methods}
The task of late aggregation methods~\cite{salehi2015preference,amer2009group,crossen2002flytrap,xiao2017fairness} is to aggregate the prediction scores (or recommendation list) of individual members as the score (or result) of the target group. That is, they first generate recommendation results for each group member, and then produce group recommendation via aggregating these individual results based on the static predefined aggregation strategies. In this category, three kinds of aggregation strategies are commonly utilized~\cite{amer2009group,salehi2015preference,quintarelli2016recommending,baltrunas2010group}, i.e., average satisfaction, least misery and maximum pleasure. For example, average satisfaction exploits the average score of all group members as the prediction score of the group by assuming each group member is equally important to the group decision making process. Least misery treats the minimum score of individual members as the prediction of the group, where the least satisfied group member plays a key role in forming the group's final decision. However, these predefined aggregation strategies are heuristic and unstable (as shown in~\cite{de2014comparison}, none of them achieves the best performance on all the datasets), only sub-optimal recommendation results are reached. 

\subsubsection{Early Aggregation methods}
Early aggregation methods are also known as preference aggregation-based methods~\cite{liu2012exploring,vinh2019interact,cao2018attentive,yuan2014generative,yin_overcoming_2020} that aim at aggregating preferences of individual group members as the profile of a group. Compared with late aggregation methods, these methods first aggregate the preference (or representation) of group members, and then make group recommendations (or produce prediction scores) accordingly. For example, \cite{liu2012exploring,yuan2014generative,gorla2013probabilistic} studied group recommendation by developing probabilistic generative models, and aggregated both the group members' individual preferences and their impacts in the group to make recommendations. Although their works treat users differently by assuming they may have different contributions to the group, both models develop the same probability distribution for each user, which is infeasible in real-world cases. To address this problem, Cao et al.~\cite{cao2018attentive} proposed an attention-based neural network to aggregate individual representation (or profile) dynamically. Vinh Tran et al.~\cite{vinh2019interact} further captured the fine-grained interactions between group members via the sub-attention networks. But their work did not consider the sparsity issue of the group-item interactions, and has limited capability in dealing with occasional group recommendations. Yin et al.~\cite{yin2019social} incorporated the social influence to \ac{OGR}, and proposed a social influence-enhanced group recommender. But the interactions among group members are ignored. To address the above challenges, Guo et al.~\cite{guo2020group} developed a group self-attention neural network to model the social influence and interactions among group members simultaneously.
He et al.~\cite{he_GAME2020} modeled the interactions among groups, users, and items with an interaction graph, and then learned their multi-view embeddings. That is, learning embeddings of groups, users, and items from their interacted counterparts to improve recommendations for occasional groups. But this work~\cite{he_GAME2020} only considers user-group interactions, group-item interactions, and user-item interactions, while the group-group correlation (i.e., group similarity) and the social homophily in the user-user interaction network are not studied. Sankar et al.~\cite{sankar_groupIM2020} proposed two data-driven strategies to investigate the preference covariance across individuals in the same group and the contextual relevance of users’ individual preferences to each group.
However, the captured preference covariance is different from our group-level similarity, which refers to the overlapping relationship among groups and can enhance the representation learning of groups by considering the groups that have common members with them.

\textbf{Differences}: Our hierarchical solution has significant differences from these existing studies. First, compared to social-interest based group recommendation methods (e.g., SIGR~\cite{yin2019social} and GroupSA~\cite{guo2020group}), we propose a different way to alleviate the sparsity issue of user-item interactions, that is, a \ac{GNN}-based user embedding network is developed to enhance the learning of users' preferences from their neighbors. Second, compared to recent neural network-based group recommendation methods (e.g., AGREE~\cite{cao2018attentive}, SIGR~\cite{yin2019social}, GroupSA~\cite{guo2020group}, GAME~\cite{he_GAME2020} and GroupIM~\cite{sankar_groupIM2020}), we further investigate the overlooked 
group similarity to alleviate the group data sparsity by treating the task of learning groups' preferences as embedding hyperedges in a hypergraph, where a weighted preference aggregation strategy is developed to consider the overlapping relationships among groups as well as the specific users who connect the groups.

\begin{figure}
    \centering
    \includegraphics[width=9cm]{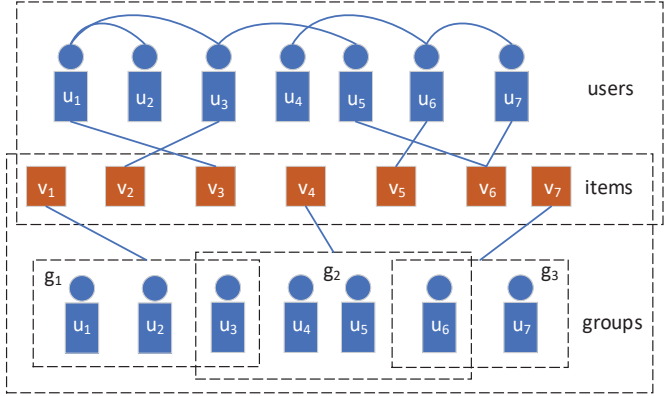}
    \caption{Illustration of the input data of our \ac{OGR} task, which includes user-item interactions, group-item interactions, user-user interactions and group-level overlapping relationships.}
    \label{fig:OGR_input}
\end{figure}

\section{Methodologies}
This section first gives an overview of our hierarchical group recommendation network, and then describes each of its components in detail.

\subsection{Preliminaries}
In this work, we use bold lowercase letters (e.g., $\boldsymbol{x}$) to represent vectors (all vectors are in column forms if not specified), and employ bold capital letters (e.g., $\boldsymbol{X}$) to represent matrices. None-bold lowercase and capital letters (e.g., $x$ and $X$) are utilized to represent scalars. Squiggle letters (e.g., $\mathcal{X}$) are used to denote sets.

Fig.\ref{fig:OGR_input} illustrates the input data of our \ac{OGR} task. Let $\mathcal{U}=\{u_1, u_2, ..., u_j, ...u_m\}$, $\mathcal{V}=\{v_1, v_2, ..., v_h, .., v_n\}$ and $\mathcal{G}=\{g_1, g_2, ..., g_t, ..., g_k\}$ be the sets of users, items and groups, and $m$, $n$ and $k$ denote the numbers of users, items and groups in the three sets respectively. The $t$-th group $g_t \in \mathcal{G}$ consists of a set of users $\mathcal{G}(t) =\{u_1, u_2, ..., u_j, ..., u_l\}$, where $u_j\in \mathcal{U}$, $l$ is the size of $g_t$, and $\mathcal{G}(t)$ is the user set of $g_t$. Each user/group interacts with different items, which indicate their preferences. Besides, users can build social connections with others, and groups have overlapping relationships with others by sharing common group members. Totally, there are four kinds of interactions among
$\mathcal{U}, \mathcal{V}$ and $\mathcal{G}$, namely user-user interactions, user-item interactions, group-item interactions and group-level overlapping relationships, which are respectively denoted by
$\boldsymbol{R}^S=[r^S_{j,j'}]^{m\times m}$, $\boldsymbol{R}^U=[r^U_{j,h}]^{m\times n}$, $\boldsymbol{R}^G=[r^G_{t,h}]^{k\times n}$ and 
$\boldsymbol{R}^H=[r^H_{t,t}]^{k\times k}$
respectively.
We use $r=1$ to indicate observed interactions, and $r=0$ for unobserved ones. 

Take $g_1$ as an illustrated example (as shown in Fig.\ref{fig:OGR_input}). Suppose $g_1$ is a target group composed of three group members $u_1, u_2$ and $u_3$. Our goal is to generate a ranked list of items that $g_1$ is likely to interact with. As $g_1$ is formed occasionally, there are limited group-item interactions, and directly learning the representation of $g_1$ (i.e., the preference of $g_1$) is not feasible. Hence, we focus on designing \ac{GNN}-based models to alleviate this sparsity issue by leveraging the user-item interactions (e.g., $(u_1, v_3), (u_2, v_1)$), social connections (e.g., $(u_1, u_2), (u_1, u_3)$) and similar groups that have common members with $g_1$ (i.e., $g_2$).
The formal definition of the group recommendation task is as follows:

\textbf{Input:} Users $\mathcal{U}$, items $\mathcal{V}$, groups $\mathcal{G}$ as well as
user-user, user-item, group-item interactions, and group-level overlapping relationships respectively denoted by $\boldsymbol{R}^S$, $\boldsymbol{R}^U$, $\boldsymbol{R}^G$ and $\boldsymbol{R}^H$.

\textbf{Output:} A function that maps an item to a real-valued score which indicates its probability of being consumed by the target group: $f_t: \mathcal{V}\to \mathbb{R}$.

\begin{figure}
    \centering
    \includegraphics[width=10cm]{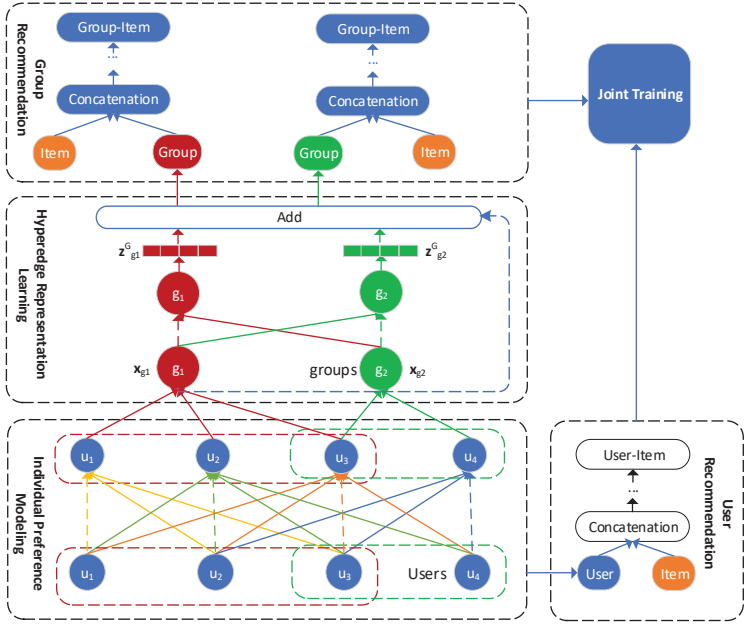}
    \caption{Overview of the architecture of \ac{HyperGroup}.}
    \label{fig:overview}
\end{figure}

\subsection{Overview of \ac{HyperGroup}}
In this work, we propose the recommender \ac{HyperGroup} for \ac{OGR}, which is designed to address the data sparsity issue in group representation learning problem with the power of a hierarchical graph neural network.

\textbf{Motivation.} Due to the sparsity nature of occasional groups, it is not straightforward to model the group preferences. The highly limited group-item and user-item interactions of \ac{OGs} make the recommendation task most challenging. An effective way to alleviate this sparsity issue is to enhance the preferences of both users and groups by leveraging the preferences of their connected neighbors, which falls into the paradigm of \ac{GNN}-based methods~\cite{Qiu2020Exploiting,hu-etal-2020-graph,wang_origin_2019}, that is, enhance users' preferences by exploiting their social connections, and enhance groups' preferences by exploiting groups that have common group members with them.
Intuitively, two groups with group members tend to have similar overall preferences, and the more common members two groups share, the higher their similarity will be. In this work, we propose to model groups as hyperedges in a hypergraph rather than treating a group as a node to build a group graph, since not only the group-level overlapping relationship, but also the specific users shared by two groups have the great potential to enhance the group preference learning. In other words, we care about not only the number of the common group members between two groups, but also who the common members are. Based on this intuition, an innovative preference aggregation strategy for hypergraph is further developed (as shown in Eq.(\ref{hyper_aggregator}), see Section \ref{group_representation}).

Fig. \ref{fig:overview} shows the architecture of our \ac{HyperGroup}, which consists of three components: \ac{IPM}, \ac{HRL} and joint model optimization.
\ac{IPM} is a \ac{GNN}-based graph embedding module that is designed to exploit users' social connections, inspired by the social principle of homophily~\cite{khanam2020homophily}, to alleviate the sparsity issue of user-item interactions,
where each group member's personal preference is enhanced by their friends' preferences and then is fed into the second component \ac{HRL} to provide foundations for the group representation learning (see Section \ref{personal}). \ac{HRL} is developed to exploit the group similarity based on common group members to learn a more accurate group representation by modeling groups as hyperedges, and the task of learning group representations is then transformed into the task of learning hyperedge embeddings (see Section \ref{group_representation}).
Finally, two joint training strategies are proposed to simultaneously optimize \ac{IPM} and \ac{HRL} in the same training process (see Section \ref{model_optimization}).

\subsection{Individual Preference Modeling\label{personal}}

Due to the rare interaction records of \ac{OGs}, directly learning their preferences by treating them as pseudo users is infeasible. Therefore, motivated by recent studies~\cite{cao2018attentive,yin2019social,guo2020group}, we
devise a paradigm that learns each group member's individual preference and then aggregates them as our initial group representation. However, due to the sparsity of individual activity data, the learned individual preference may be not accurate. To address this challenge, motivated by the social principle of homophily~\cite{khanam2020homophily}, we develop a social-enhanced individual preference modeling method in the \ac{IPM} component.

As social animals, users turn to their friends for recommendations~\cite{fan2019graph,chen_social2020,wang2019Social}, and also share many common preferences with friends. Thus, users' preferences can be summarized from both themselves and their direct neighbors (i.e., friends) in a social network~\cite{mcpherson2001birds,guo2016social}.
Specifically, to learn users' preferences from their social neighbors (denoted as $\boldsymbol{R}^S$), we treat users as nodes in a large social graph and develop a \ac{GNN}-based node embedding module, where the preferences of their neighbors and their own are simultaneously considered.
It is worth mentioning that, compared with the spectral graph convolutional network~\cite{kipf2016semi}, we build our model upon an information aggregation-based network~\cite{hamilton2017inductive,Rossi2018_deep}, which is an inductive representation learning approach that bypasses the need for the entire graph’s node adjacency matrix to operate, thus being space-efficient when handling large-scale datasets for recommendation.

\begin{algorithm}
\caption{Individual embedding generation algorithm}
\label{alg:graphsage}
\begin{algorithmic}[1]
\REQUIRE User-user interactions $\boldsymbol{R}^S$; input features $\{\boldsymbol{x}_u, \forall u \in \mathcal{U}\}$; depth $K$; weight matrices $\boldsymbol{W}^i, \forall i \in \{1, ..., K\}$; non-linearity $\sigma$; differentiable aggregator functions AGGEGATE$_i$, $\forall i \in \{1, ..., K\}$; neighborhood function $\mathcal{N}: u\to 2^\mathcal{U}$
\ENSURE Vector representations $\boldsymbol{z}_u$ for all $u\in \mathcal{U}$
\STATE $\boldsymbol{h}_u^0 \leftarrow \boldsymbol{x}_u, \forall u\in \mathcal{U}$;
\FOR{$i=1...K$}
\FOR{$u\in \mathcal{U}$}
\STATE $\boldsymbol{h}_{\mathcal{N}(u)}^i \leftarrow \text{AGGREGATE}_i(\{\boldsymbol{h}_{u'}^{i-1}, \forall u' \in \mathcal{N}(u)\})$;
\STATE $\boldsymbol{h}_u^i\leftarrow \sigma(\boldsymbol{W}^i\cdot concat(\boldsymbol{h}_u^{i-1}, \boldsymbol{h}_{\mathcal{N}(u)}^i))$;
\ENDFOR
\STATE $\boldsymbol{h}_u^i \leftarrow \boldsymbol{h}_u^i/||\boldsymbol{h}_u^i||_2, \forall u \in \mathcal{U}$;
\ENDFOR
\STATE $\boldsymbol{z}_u\leftarrow \boldsymbol{h}_u^K, \forall u\in \mathcal{U}$;
\end{algorithmic}
\end{algorithm}
 
Algorithm \ref{alg:graphsage} shows the details of our \ac{GNN}-based individual preference modeling component, which takes a social graph $\boldsymbol{R}^{S}$ with its node/user features $\boldsymbol{x}_{u}, u\in \mathcal{U}$ as the input. 
We adopt the node embedding method node2vec~\cite{grover2016node2vec} to obtain the initialized node/user features, as it can well balance the embedding quality and computation cost.
In the outer loop of this algorithm, each user $u\in \mathcal{U}$ first aggregates the representations of the nodes in its immediate neighborhood, $\{\boldsymbol{h}_{u'}^{i-1}, \forall u' \in \mathcal{N}(u)\}$, into a single vector $\boldsymbol{h}_{\mathcal{N}(u)}^{i-1}$, where $\mathcal{N}(u)$ is the sampled neighbors of $u$ with a fixed size $S$, $i$ denotes the $i$-th iteration or the $i$-th layer of \ac{GNN} and $\boldsymbol{h}^i$ denotes a node's representation at this iteration or layer. 
For $i =0$, we let $\boldsymbol{h}^{0}_{\mathcal{N}(u)} = \boldsymbol{x}_u$. 
Note that a node's representation $\boldsymbol{h}^{i}_{u}$ at $i$-th iteration depends on
both its own representation $\boldsymbol{h}^{i-1}_u$ and the aggregated neighborhood vector representation $\boldsymbol{h}^{i}_{\mathcal{N}(u)}$ generated at the $i$-$1$-th layer. We adopt the concatenation operation to combine them, followed by 
a fully connected layer with nonlinear activation function $\sigma(\cdot)$ and weight matrices $\boldsymbol{W}^{i}, \forall i \in \{1,...,K\}$, which are used to propagate information between different layers. 
The final output at the $K$-th layer is denoted as $\boldsymbol{z}_u = \boldsymbol{h}_u^K, \forall u \in \mathcal{U}$ for convenience, which encodes $u$'s preferences.

The aggregation function that aims at aggregating neighbor representations ( denoted by AGGREGATE$(\cdot)$ in Algorithm \ref{alg:graphsage}) can be done by a variety of aggregator architectures~\cite{hamilton2017inductive} (e.g., mean aggregator, max-pooling aggregator and LSTM aggregator). In this work, we simply take the mean aggregator as our aggregation function, where the element-wise mean operation is applied to aggregate information across the neighbor set:
\begin{align}
    \text{AGGREGATE}_i(u) = \text{MEAN}(\boldsymbol{h}_{u'}^{i}, \forall u' \in \{\mathcal{N}(u)\})
\end{align}

Then, these learned individual embeddings within an occasional group are further aggregated~\cite{liu2017attention} to produce the group representation in the individual preference space. The embedding of group $g$ is denoted as:
% \vspace{-0.2cm}
\begin{align}
    \boldsymbol{x}^G_g =
    \sum_{\forall u \in \mathcal{G}(g)} \alpha_{u}  \boldsymbol{emb}_u,
    \label{hyper_aggregator}
\end{align}
where $\boldsymbol{emb}_u=\boldsymbol{z}_u + \boldsymbol{z}'_u$, $\boldsymbol{z}_u$ is the learned individual representation of group member $u$, $\boldsymbol{z}'_u$ is $u$'s embedding in latent feature space, and $\alpha_{u}$ denotes the importance of user embedding. As this work is not focused on designing preference aggregation strategies, the simple average operation is utilized (i.e., $\alpha_{u}= 1/|\mathcal{G}(g)|$). The output of \ac{IPM} ($\boldsymbol{x}^G_g$) is then passed to the \ac{HRL} component as the initial representations of the corresponding group, which will be further optimized and learned in the higher-layer of our framework.

Compared to existing transductive individual preference learning methods that are based on matrix factorization
% Unlike most transductive node embedding methods that are based on matrix factorization
~\cite{xu2017embedding, wang2017community}, \ac{IPM} leverages node features to learn an embedding function that generalizes to unseen nodes, where the topological structure of each node's neighborhood and the distribution of node features in the neighborhood are simultaneously learned. Moreover, rather than training a distinct embedding vector for each node, we train $K$ aggregation functions to aggregate feature information from the local neighborhood. Each aggregator function aggregates information from a different number of hops away from a given node~\cite{hamilton2017inductive,petar_graph_2018}. Note that, the number of hops is equal to the number of aggregators (both denoted as $K$).

\begin{algorithm}[t]
\caption{Hyperedge embedding generation process}
\label{alg:hyperedge}
% \footnotesize
\begin{algorithmic}[1]
\REQUIRE Hypergraph $\mathcal{G}^G(\mathcal{U},\mathcal{G})$; learned features $\{\boldsymbol{x}^G_g, \forall g \in \mathcal{G}\}$ from \ac{IPM}; depth $K$; weight matrices $\boldsymbol{W}^i, \forall i \in \{1, 2, ..., K\}$; nonlinear activation function $\sigma$; aggregator functions $\text{AGGEGATE}_i^G$, $\forall i \in \{1, 2, ..., K\}$; neighborhood function $\mathcal{N}: g\to 2^\mathcal{G}$
\ENSURE Hyperedge embeddings $\boldsymbol{z}_g^G$ for all $g\in \mathcal{G}$
\STATE $\boldsymbol{m}_g^0 \leftarrow \boldsymbol{x}^G_g, \forall g\in \mathcal{G}$;
\FOR{$i=1...K$}
\FOR{$g\in \mathcal{G}$}
\STATE $\boldsymbol{m}_{\mathcal{N}(g)}^i \leftarrow \text{AGGREGATE}_i^G(\{\boldsymbol{m}_{g'}^{i-1} + \boldsymbol{l}_{g,g'}^i, \forall g' \in \mathcal{N}(g)\})$;
\STATE $\boldsymbol{m}_g^i\leftarrow \sigma(\boldsymbol{W}^i\cdot concat(\boldsymbol{m}_g^{i-1}, \boldsymbol{m}_{N(g)}^i))$;
\ENDFOR
\STATE $\boldsymbol{m}_g^i \leftarrow \boldsymbol{m}_g^i/||\boldsymbol{m}_g^i||_2, \forall g \in \mathcal{G}$;
\ENDFOR
\STATE $\boldsymbol{z}^G_g\leftarrow \boldsymbol{m}_g^K, \forall g\in \mathcal{G}$;
\end{algorithmic}
\end{algorithm}

\subsection{Hyperedge Embedding-based Group Representation Learning\label{group_representation}}

Simply aggregating group members' individual preferences as the preference of a group would miss the intrinsic group-level preferences which may be different from all individuals' preferences within the group. Moreover, as a user may belong to multiple groups at the same time, the groups that have common members should have similar group-level preferences.
To quantify the similarity between two groups, the common users between two groups are further exploited to model the group-level similarity. 
What matters is not only the number of the common users, but also who the common users are, as different users have different influences on the group decision making. For example, suppose groups $g_1, g_2$ share a common user $u_1$, and groups $g_2, g_3$ share a common user $u_2$. If we know $u_1$ is more influential than $u_2$, we can infer that, the similarity between $g_2$ and $g_1$ is higher than that between $g_2$ and $g_3$. On this basis, let us further assume that $g_2$ has sufficient historical data and $g_1$ is a cold-start group, then $g_2$'s preferences would provide important signals for $g_1$'s preferences, thus further alleviate the data sparsity issue and produce more effective group representations for \ac{OGR}.

To capture the group-level preferences and further exploit the group similarity to enhance group preference learning, we innovatively introduce a hypergraph~\cite{berge1984hypergraphs,wang2020Next,bai2019hypergraph} to model groups rather than treating a group as a node to build a group graph. That method would fail to capture who the common users are. In a hypergraph, each group is treated as a hyperedge that connects all users in that group, and two hyperedges are incident if they share at least one common member.

Then, the task of learning group representations is transformed into embedding hyperedges in a hypergraph.
To integrate the group similarity based on common members in the learning process of hyperedge embedding, we devise a GNN-based hyperedge embedding model, called \ac{HRL}.
As \ac{OGs} tend to be formed by chance~\cite{guo2020group, yin2019social}, \ac{HRL} also adopts an inductive graph embedding method~\cite{hamilton2017inductive,bai2019unsupervised} as its building block to generate embeddings for hyperedges, where a weighted feature aggregation scheme is proposed to account for the similarity between two groups.

Formally, given $k$ groups $\mathcal{G}=\{g_1, g_2, ..., g_t, ..., g_k\}$ defined over the user set $\mathcal{U}$, where $g_t = \mathcal{G}(t) = \{u_1, u_2, ..., u_j, ...u_l\}$ consists of a set of users, we first construct a hypergraph $\mathcal{G}^G =(\mathcal{U}, \mathcal{G}$), where $\mathcal{G}$ is the collection of hyperedges/groups over the nodes/users $\mathcal{U}$. Let $\boldsymbol{H}\in \{0,1\}^{|\mathcal{G}|\times|\mathcal{U}|}$ represent the incidence matrix of $\mathcal{G}^G$ with $\boldsymbol{H}(g,u)=1$ if $u\in \mathcal{G}(g)$ else 0. The degree $d(u)$ of a vertex $u$ is defined as the number of hyperedges associated with $u$, i.e., $d(u)= \sum_{g\in \mathcal{G}} \boldsymbol{H}(g,u)$. To generalize to an unobserved group, a hyperedge embedding generator will be learned and its basic idea is to aggregate feature information from its incident hyperedges as its embedding (i.e., its preference encoding).

Algorithm \ref{alg:hyperedge} describes the hyperedge embedding generation process, where the constructed hypergraph, $\mathcal{G}^G (\mathcal{U},\mathcal{G})$, and the learned features $\boldsymbol{x}^G_g, \forall g \in \mathcal{G}$ from \ac{IPM} are provided as input. 
In the higher-layer of our network, $\boldsymbol{x}^G_g, \forall g \in \mathcal{G}$ serves as the initial representation of group $g$, which will be further optimized. 
At each iteration or layer $i$, a hyperedge $g\in \mathcal{G}$ aggregates features from its immediate local neighbors (i.e., its incident hyperedges). Specifically, a hyperedge $g \in \mathcal{G}$ aggregates the embeddings of its incident hyperedges, $\{\boldsymbol{m}_{g'}^{i-1}, \forall g' \in \mathcal{N}(g)\}$, into a single vector $\boldsymbol{m}_{\mathcal{N}(g)}^{i}$ (the size of $\mathcal{N}(g)$ is also set as $S$), where $\boldsymbol{m}^{i-1}_{g'}$ denotes the embedding of $g'$ generated at the previous the iteration/layer $i-1$. After that, the aggregated neighborhood vector is concatenated with the hyperedege's previous representation, $\boldsymbol{m}_g^{i-1}$, and followed by a fully connected layer with nonlinear activation function $\sigma(\cdot)$ and weight matrices $\boldsymbol{W}^{i}, \forall i \in \{1,...,K\}$, which are used to search depths. The representation obtained at the last layer/iteration $\boldsymbol{m}^{K}_{g}$ denotes the group $g$'s preferences, which is defined as $\boldsymbol{z}_g^G = \boldsymbol{m}_g^K, \forall g \in \mathcal{G}$. We define our hyperedge aggregator $\text{AGGREGATE}^G$ as follows:
\begin{align}
    \text{AGGREGATE}_i^G (g) =
    \sum_{\forall g' \in \mathcal{N}(g)} \alpha_{g,g'} (\boldsymbol{m}_{g'}^i + \boldsymbol{l}_{g,g'}^i) 
    \label{hyper_aggregator}
\end{align}
where $\alpha_{g,g'}$ is the aggregation weight determined by the similarity between group $g$ and $g'$.

In this work, we set $\alpha_{g,g'}$ to number of the common members between group $g$ and $g'$. By doing this, we are able to give more attention to the groups sharing more common members with $g$. In Eq.(\ref{hyper_aggregator}), $\boldsymbol{l}_{g,g'}$ denotes the representation of the set of common members between group $g$ and $g'$, which is used to distinguish the specific members who are shared by these two groups. For simplicity, MEAN aggregator is employed to compute $\boldsymbol{l}_{g,g'}$:
% \vspace{-0.1cm}
\begin{align}
    \boldsymbol{l}_{g,g'}^i=\text{MEAN}(\boldsymbol{emb}_{u}, \forall u\in \{\mathcal{G}(g) \cap \mathcal{G}(g')\}).
\end{align}
To this end, our proposed representation aggregator not only considers the number of the shared common users, but also emphasises the importance of individuals who are the shared group members.

The resulted group representation $\boldsymbol{z}^G_g$ is then added with $\boldsymbol{x}^G_g$ by the residual  operation. The final embedding of group $g$ (denoted as $\boldsymbol{emb}^G_g$) can be represented as:
\begin{align}
    \boldsymbol{emb}^G_g = w \boldsymbol{z}^G_g + (1 - w) \boldsymbol{x}^G_g,
\label{eq:residual}
\end{align}
where $w$ is a hyper-parameter controlling the contributions of the two parts.

Note that, our solution is different from existing graph neural networks ~\cite{hamilton2017inductive,bai2019unsupervised} in two aspects. First, our group representation learning network is hierarchical, which first learns the group members' personal preferences in the lower-layer (i.e., \ac{IPM}) and then infers groups' representations in the higher-layer (i.e., \ac{HRL}). Second, to exploit and integrate the group similarity based on common members in \ac{HRL}, the preference aggregation strategy (a.k.a hyperedge aggregator) that considers both the group-level preference and the personal preferences of common group members is devised.

\subsection {Model Optimization\label{model_optimization}}
Given the embeddings of the target group and item (i.e., $\boldsymbol{emb}_g^G$ and $\boldsymbol{emb}_h^V$), we feed the concatenation of them into a \ac{MLP} for preference prediction (as shown in Fig.\ref{alg:graphsage}):
\begin{align}
    & \boldsymbol{c}_1^G =[\boldsymbol{emb}_g^G \oplus \boldsymbol{emb}_h^V] \notag \\
    & \boldsymbol{c}_2^T = \sigma (\boldsymbol{W}_2 \cdot \boldsymbol{c}_1^G + \boldsymbol{b}_2) \notag \\
    & \dots \notag \\
    & \hat{r}_{g,h}^G = \boldsymbol{w}^T \cdot \boldsymbol{c}_{k-1}^G
\end{align}
where $\boldsymbol{emb}_h^V$ is the item embedding in the latent space that is learned via optimizing the following loss function (Eq.(\ref{eq:group_rank})); $\boldsymbol{W}$ and $\boldsymbol{b}$ are the weight and bias of a feed-forward network; $\hat{r}_{g,h}^G$ is the predicted preference score of group $g$ to item $v_h$.

Due to the implicit nature of the group-item interaction data, motivated by~\cite{guo2020group}, a pairwise loss function~\cite{rendle2009bpr} is employed:
\begin{align}
    L_G \!=\! \text{arg}\min_\Theta\! \sum_{(g,v_h,v_{h'})\in \mathcal{D}_G}\!-\text{ln}\sigma(\hat{r}^G_{g,h}\! -\! \hat{r}^G_{g,h'})\!+\! \lambda ||\boldsymbol{\Theta}||^2
\label{eq:group_rank}
\end{align}
where $\boldsymbol{\Theta}$ represents the set of the model parameters. To learn from this implicit feedback, we reconstruct the group-item data by assuming that groups prefer observed item $v_h$ over all other unobserved item $v_h'$. Then, the training data $\mathcal{D}_G: \mathcal{G}\times \mathcal{V}\times \mathcal{V}$ can be denoted as:
\begin{align}
    \mathcal{D}_G =\{(g, v_h, v_{h'})| v_h \in \mathcal{V}_g^+ \land v_{h'}\in \mathcal{V}\setminus \mathcal{V}_g^+\}
\end{align}
where $\mathcal{V}_g^+$ and $\mathcal{V}\setminus \mathcal{V}_g^+$ are the observed and unobserved item set w.r.t group $g$. In this work, we use $N_x$ to denote the number of sampled negative items per positive item. The meaning of $(g, v_h, v_{h'})\in \mathcal{D}_G$ is that group $g$ prefers item $v_h$ over $v_{h'}$.

As in \ac{OGs}, the available group-item interactions are extremely sparse, the learned group representations (via optimizing Eq. (\ref{eq:group_rank})) are not sufficiently accurate or reliable.
To further accelerate and enhance the group preference learning, we propose to leverage the user-item interaction data to optimize the group-item and user-item recommendation tasks simultaneously.
As shown in Fig. \ref{fig:overview}, we propose to use another MLP to model the user-item interaction data. More specifically, given the embeddings of the target user and item, we first feed them into a \ac{MLP} to calculate the personal preference score of a user to an item:
\begin{align}
    & \boldsymbol{c}^U_1 = [ \boldsymbol{emb}_u \oplus \boldsymbol{emb}_h^V] \notag \\
    & \boldsymbol{c}^U_2 = \sigma (\boldsymbol{W}_2 \cdot \boldsymbol{c}^U_1 +\boldsymbol{b}_2) \notag \\
    & ... \notag \\
    & \hat{r}_{u,h}^{U} = \boldsymbol{w}^T \cdot \boldsymbol{c}^U_{k-1}
\label{eq:user_rank}
\end{align}
where $\boldsymbol{emb}_u$ is the shared user representation/embedding that connects all user-user, user-item and group-item spaces and  data. $\hat{r}_{u,h}^{U}$ is the predicted preference score of user $u$ to item $v_h$. $\boldsymbol{emb}_h^V$ is another shared item embedding that bridges the group-item space and user-item space. As user-item interaction data is also implicit, the same pairwise loss function is utilized:
\begin{align}
     L_U \!=\! \text{arg} \min_\Theta \! \sum_{(u,v_h,v_{h'})\in \mathcal{D}_U}-\text{ln}\sigma(\hat{r}^U_{u,h} \!-\! \hat{r}^U_{u,h'})\!+\! \lambda ||\boldsymbol{\Theta}||^2
\label{final:user}
\end{align}
where $\mathcal{D}_U$ denotes the set of reconstructed user-item samples; $(u, v_h, v_{h'})$ represents user $u$ prefers observed item $v_h$ over unobserved item $v_{h'}$. Similar to Eq. (\ref{eq:group_rank}), for every positive item, $N_x$ negative items are randomly sampled. 
Technically, to integrate $L_G$ with $L_U$, we develop two model optimization approaches Two-stage Training and Joint Training.

\begin{algorithm}[t]
\caption{Two-state training method of \ac{HyperGroup}}
\label{alg:two-stage-training}
\begin{algorithmic}[1]
\REQUIRE $\mathcal{R}^S, \mathcal{R}^U, \mathcal{R}^G, \mathcal{R}^H$, number of positive samples of users $M_u$, number of positive samples of groups $M_g$, number of negative samples $N_x$;
\ENSURE Parameter set $\Theta = \{\boldsymbol{emb}_g^G, \boldsymbol{emb}_h^V, \boldsymbol{emb}_u, \boldsymbol{W}, \boldsymbol{b}\}$;
% \vspace{-0.3cm}
\WHILE{$iter \leq$ $M_u$} 
    \STATE Randomly draw $(u, v_h)$ from $\mathcal{R}^U$; 
    \STATE Randomly sample $N_x$ negative examples for $u$;
    \STATE Update the model parameters w.r.t. Eq. (\ref{final:user});
\STATE $iter = iter+1$;
\ENDWHILE
\WHILE{$iter \leq$ $M_g$} 
    \STATE Randomly draw $(g, v_h)$ from $\mathcal{R}^G$; 
    \STATE Randomly sample $N_x$ negative examples for $g$;
    \STATE Update the model parameters w.r.t. Eq. (\ref{eq:group_rank}); 
\STATE $iter = iter+1$;
\ENDWHILE
\end{algorithmic}
\end{algorithm}

\textbf{Two-stage Training.} 
In this strategy, we first optimize $L_U$ by the user-item interaction data to learn the representations of users and items in the user-item space, and then take item latent features as the latent vector of items in the group-item recommendation task (as shown in Algorithm~\ref{alg:two-stage-training}). In the second stage, the parameters will be fine-tuned by optimizing $L_G$ with the group-item interactions. In both of these two training stages, the Stochastic Gradient Descent (SGD) algorithm is adopted, and at each gradient step, a positive user-item sample ($u, v_h$) (or group-item example ($g, v_h$)) and $N_x$ negative corresponding samples ($u, v_{h'}$) (or ($g, v_{h'}$)) are randomly selected for training.

\begin{algorithm}[t]
\caption{Joint training method of \ac{HyperGroup}}
\label{alg:joint}
\begin{algorithmic}[1]
\REQUIRE $\mathcal{R}^S, \mathcal{R}^U, \mathcal{R}^G, \mathcal{R}^H$, number of positive samples of users $M_u$, number of positive samples of groups $M_g$, number of negative samples $N_x$;
\ENSURE Parameter set $\Theta = \{\boldsymbol{emb}_g^G, \boldsymbol{emb}_h^V, \boldsymbol{emb}_u, \boldsymbol{W}, \boldsymbol{b}\}$;
% \vspace{-0.3cm}
\WHILE{$iter \leq$ ($M_u + M_g$)} 
    \STATE Randomly draw $(u, v_h)$ from $\mathcal{R}^U$ and sample $N_x$ negative examples for $u$, and update the model parameters w.r.t. Eq. (\ref{final:user});
   
    \STATE Randomly draw $(g, v_h)$ from $\mathcal{R}^G$ and sample $N_x$ negative examples for $g$, and update the model parameters w.r.t. Eq. (\ref{eq:group_rank}); 
\STATE $iter = iter+1$;
\ENDWHILE
\end{algorithmic}
\end{algorithm}

\textbf{Joint Training.} In this strategy, we jointly train $L_G$ and $L_U$ on all the group-item and user-item interactions (as shown in Algorithm~\ref{alg:joint}), and the loss function is actually changed to the following equation:
\begin{align}
    L(\boldsymbol{\Theta}) = L_G (\boldsymbol{\Theta}) + L_U (\boldsymbol{\Theta}).
\end{align}
All the parameters (denoted by $\boldsymbol{\Theta}$) are learned by a standard Stochastic Gradient Descent (SGD), and at each gradient step, we first randomly draw a positive user-item sample ($u, v_h$) and a positive group-item example ($g, v_h$)) from the user-item set and group-item set respectively, and then draw $N_x$ negative corresponding samples ($u, v_{h'}$) and ($g, v_{h'}$) to update the gradients.
\section{Experimental Setup}

In this section, we first introduce the research questions that we aim to answer in experiments, and then describe the datasets, evaluation methods and baselines utilized in this work.

\subsection{Research Questions}
We conduct extensive experiments on two real-world datasets to answer the following research questions.
\begin{itemize}
\item[\textbf{RQ1}] How does our proposed \ac{HyperGroup} approach perform compared with state-of-the-art group recommendation methods?
\item[\textbf{RQ2}] How do the three components of \ac{HyperGroup}, i.e., \acf{IPM}, \acf{HRL}, and the joint training method contribute to the performance of \ac{HyperGroup}? How do our proposed model optimization approaches perform on heterogeneous interaction data?
\item[\textbf{RQ3}] How do the hyper-parameters affect the performance of \ac{HyperGroup}?
\item[\textbf{RQ4}] How is the training efficiency and scalability of \ac{HyperGroup} when processing large-scale data?
\end{itemize}

\begin{table}
\caption{Statistics of the datasets.}
\centering
% \footnotesize
\begin{tabular}{ccc}
\toprule
\textbf{Statistics}	& \textbf{Yelp}	& \textbf{Douban-Event}\\
\midrule
\# Users & 34,504 & 29,181\\
\# Groups & 24,103 & 17,826\\
\# Items/Events		& 22,611 & 46,097\\
Avg. group size & 4.45 & 4.84 \\
Avg. \# interactions per group	& 1.12	& 1.47\\
Avg. \# friends per user &20.77 &40.86 \\
Avg. \# interactions per user  & 13.98 &25.22\\
\bottomrule
\label{tab_statistic}
\end{tabular}
\end{table}

\subsection{Datasets}

To evaluate the performance of our \ac{HyperGroup} method, we conduct experiments on two large-scale real-world datasets Yelp\footnote{www.yelp.com} and Douban-Event\footnote{www.douban.com/location/world/} that are exclusively published for \ac{OGR} by Yin et al.~\cite{yin2019social}. %add the nature of two datasets
Yelp is a famous online social network that connects people with local businesses (e.g., restaurants and home services), where users can publish their reviews about these businesses and create social connections. The published dataset only focuses on the restaurants located in Los Angeles and every record in it contains a user, a timestamp and a business, which indicates the user visited the restaurant at that time. Douban-Event is one of the largest online event-based social networks in China that helps people publish and participate in social events. In this dataset, the user's event attendance list and friend list, as well as the event's time and venue were collected.

As the raw data of these two datasets does not contain any explicit group information, Yin et al.~\cite{yin2019social} extracted implicit groups by the following strategy:
if a set of users who are connected in the social network attend the same event or visit the same restaurant at the same time, they are defined as the members of a group, and the group activities are the common activities of these users.
The resulted Yelp data has 34,504 users, 24,103 groups, and 22,611 items for training and testing.
For the Douban-Event data, to reduce the data size, we follow the data used in~\cite{guo2020group}, which is generated by randomly keeping 29,181 users, 17,826 groups and 46,097 items. 
The statistics of these two datasets are shown in Table \ref{tab_statistic}, from which we have the following observation, that is, compared with user-item interactions, the group-item interactions are much sparser. For example, in Yelp a user has 20.77 interactions on average, while a group has only 1.12 interactions. The second observation is that the user-item interaction data is also quite sparse. The densities of the user-item interaction matrices for Yelp and Douban-Event are 0.051\% and 0.057\%, respectively.

Note that, as the other two datasets CAMRa2011\footnote{http://2011.camrachallenge.com/2011} and Movielens-Group~\cite{yuan2014generative} have either persistent groups or randomly generated groups, and none of them contain the social network information, they are not suitable to evaluate our solution. We do not conduct experiments on these two datasets.

\subsection{Evaluation Protocols}

We randomly split each dataset into training, validation and test sets with the ratio of 80\%, 10\% and 10\% respectively.
To fully evaluate our proposed method, we do not follow the evaluation protocol proposed in ~\cite{cao2018attentive,guo2020group}, which only randomly selects 100 items that have never been interacted by the tested group as the candidate set to be ranked. Instead, we evaluate all the comparison methods by testing their ability to rank all items for each tested group, and report their performance in recommending Top-$N$ items.
The evaluation metrics \ac{HR} and \ac{NDCG}~\cite{he2015trirank} are adopted in our experiments, where \ac{HR} measures how many candidate items are ranked within the Top-$N$ list, while \ac{NDCG} accounts for the position of the hit by assigning higher score to hit at top positions. 

More specifically, for each group-item interaction ($g,v$) in the test set, we first compute the ranking score for item $v$ and all candidate items. And then, we pick $N$ items with the highest ranking scores as the Top-$N$ recommendation list. If item $v$ appears in this list, we have a hit. Otherwise, we have a miss. The formal definition of \ac{HR}~\cite{he2015trirank} is written as follows:
\begin{equation}
    \ac{HR}@N = \frac{\#hit@N}{|\mathcal{D}_{test}|}
\end{equation}
where $\#hit@N$ denotes the number of hits in the test set, and $|\mathcal{D}_{test}|$ is the total number of the test cases. 

The metric \ac{NDCG} \cite{he2015trirank} is defined as:
\begin{equation}
    \ac{NDCG}@N = Z_N \sum_{i=1}^N \frac{2^{r_i}-1}{log_2(i+1)}
\end{equation}
where $Z_N$ is the normalizer\footnote{We set $Z_N=\text{log}(2)$, as we use the binary relevance of item. } to ensure that the perfect ranking has a value of 1; $r_i$ is the graded relevance of item at position $i$. We use the simple binary relevance in this work, that is, if the item at position $i$ is the ground-truth item, $r_i =1 $; otherwise $r_i =0$.

\subsection{Baseline Methods\label{baselines}}
We compare \ac{HyperGroup} with the following baseline methods.

\begin{itemize}
    \item \textbf{Pop~\cite{cremonesi2010performance}.} This is a popularity-based recommendation method, which recommends the most popular items in the training set.
    \item \textbf{NCF ~\cite{he2017neural}.} This method is developed for individual users. We utilize this method for \ac{OGR} by treating groups as virtual users.
    \item \textbf{BPR-MF ~\cite{rendle2009bpr}.} This is a traditional collaborative filtering-based method exploiting the pairwise loss as the optimization objection for recommending items to individual users. Same as NCF, we used it for \ac{OGR} via assuming groups are virtual users.
    \item \textbf{PIT~\cite{liu2012exploring}.} This is a probabilistic model devised for \ac{OGR}, which extends the author topic model~\cite{Rosen_the_author2004} proposed for document-authorship analysis by treating a group of users as the authors of a document and the interacted items as the words of the document. In this method, a personal impact parameter is introduced to model the representativeness of each member to a group.
    % but it ignores the fact that the representative of a user will be different across different topics. 
    \item \textbf{COM ~\cite{yuan2014generative}.} This is another topic model-based approach proposed for \ac{OGR}, but different from PIT that only considers group members' own topic preferences to select items, it considers both members’ topic-dependent influences and group behaviors.
    \item \textbf{AGREE~\cite{cao2018attentive}.} This is the first work that employs a neural attention network to learn the dynamic aggregation strategy for \ac{OGR}.
    \item \textbf{SIGR~\cite{yin2019social}.} This work develops a deep social influence learning framework to exploit both global and local social network structures to learn the social influence or weight of each group member in the group decision making. This is the first work to focus on the data sparsity issues of \ac{OGR}.
    \item \textbf{GroupSA~\cite{guo2020group}.} This is the state-of-the-art group recommendation method proposed for \ac{OGR}, where the self-attention mechanism is utilized to learn the group preference aggregation strategies.
    \item \textbf{GroupIM~\cite{sankar_groupIM2020}.} This is another state-of-the-art group recommendation method developed for \ac{OGR}, which leverages two data-driven strategies to investigate the preference covariance across individuals in the same group and the contextual relevance of users’ individual preferences to each group. However, the captured preference covariance is different from our group-level similarity, which refers to the overlapping relationship among groups and can enhance groups’ preferences via exploiting groups that have common group members with them. 
\end{itemize}

\subsection{Implementation Details}
We implement \ac{HyperGroup} based on Pytorch accelerated by NVIDIA RTX 2080 Ti GPU. In experiments, we first initialize the parameters using the Glorot initialization method~\cite{glorot2010understanding}, and then use the Adam optimizer~\cite{kingma2014adam} to optimize our loss function, where the mini-batch size is set to 256, and the initial learning rate is set to 0.0001. For hyper-parameters, the number of negative samples ($N_x$) per positive sample is searched within $\{1, 2, 3, 4, 5\}$; the dimensions of the network features, the embeddings of user, group and item are all set to 128; the number of latent layers is set as $K=1$ for \ac{IPM} and $K=2$ for \ac{HRL}; the number of sampled neighbors (denoted as $S$) for \ac{IPM} and \ac{HRL} are both searched within $\{1, 2, 3, 4, 5\}$; the hyper-parameter $w$ that determines the importance of the residual connection is searched within [0.1-0.9] with a step size of 0.1. The details of tuning the hyper-parameters are shown in Section~\ref{hyperparameters}. To avoid over-fitting, the dropout regularization method~\cite{nitish2014_dropout} with drop ratio 0.1 is utilized for both datasets. If not specified, 
all the reported experimental results of our methods are achieved with a Two-stage Training strategy.

For the settings of baseline methods, we tune the following hyper-parameters that are reported as important factors in their publications to obtain optimal performance, and let the others as the default setting (both datasets are applied): 1) For NCF, we set the learning rate = [0.0005, 0.0001, 0.00005], negative samples = 3, and dropout ratio =0.1. 2) For BPR-MF, we set the factor number = 30, and sampled triples = $\sqrt{MaxUserID} \times 100$. 3) For PIT and COM, we tune the number of topics and achieve the best result when topic number $=250$.
4) For AGREE, we set the learning rate = $[0.005, 0.001, 0.0005]$, and negative samples = 1. For fair comparisons, in all the ranking-based methods developed for \ac{OGR} (i.e., AGREE, SIGR, and GroupSA), the number of negative samples per positive sample is set as 1 (as the setting in \ac{HyperGroup}). 5) For SIGR, we set the importance controller $\eta = 0.5$ and $1/\rho^2_S = 0.05$. 6) For GroupSA, we set the self-attention layer as 2, the number of items (or users) utilized in the item aggregation (or social aggregation) as 4. 7) For GroupIM, we set the layer size = 64, and negative users sampled per group = 5. 

Note that, all the baselines are trained end-to-end and the neural network-based methods (i.e., NCF, AGREE, SIGR, GroupSA, GroupIM and \ac{HyperGroup}) are optimized with no pre-training.
\begin{table*}[ht]
  \caption{Top-$N$ Recommendation performance on Yelp and Douban-Event via evaluating on all items.}
  \label{tab:results_all}
  \centering
%   \scriptsize
%   \small
  \begin{tabular}{|c|c|c|c|c|c|c|c|c|}
    \hline
    \multicolumn{9}{|c|}{Overall Performance Comparison} \\
    \hline
    \multirow{3}*{Methods} &\multicolumn{4}{c|}{Yelp} &\multicolumn{4}{c|}{Douban-Event} \\
    \cline{2-9}
    & \multicolumn{2}{c|}{$N$=5} &\multicolumn{2}{c|}{$N$=10}& \multicolumn{2}{c|}{$N$=5} &\multicolumn{2}{c|}{$N$=10} \\
    \cline{2-9}
    &HR & NDCG & HR& NDCG &HR & NDCG & HR& NDCG\\
    \hline
    Pop &0.0117 &0.0076 &0.0201 &0.0103
    &0.0031 &0.0017 &0.0046 &0.0022\\
     NCF &0.0110 &0.0074 &0.0193 &0.0100
     &0.0041 &0.0024 &0.0061 &0.0030\\
     BPR-MF &0.0026 &0.0056&0.0022 &0.0078
     &0.0009 &0.0017 &0.0007 &0.0023\\
     \hline
     PIT &0.0128 &0.0076 &0.0258 &0.0117
     &0.0079 &0.0043 &0.0190 &0.0075\\
    COM &0.0481 & 0.0313& 0.0812 &0.0420
    &0.0103& 0.0053 &0.0214 &0.0089\\
    \hline
    AGREE &0.0569 & 0.0389 & 0.0896 &0.0495
    &0.0122& 0.0073 &0.0255 &0.0116\\
  SIGR &0.1085 &0.0738&0.1499 &0.0871
  &0.0200 &0.0114 &0.0345 &0.0162\\
  GroupSA &0.1211 &0.0843 &0.1680 &0.0992 
&0.0212 &0.0137 &0.0382 &0.0191\\
  GroupIM &0.1312 &0.1033 &0.1493 &0.1090
&0.0511 &0.0358 &0.0669 &0.0406\\
    \hline
    \ac{HyperGroup} &\textbf{0.4827} &\textbf{0.3973} &\textbf{0.5598}&\textbf{0.4223}
    &\textbf{0.0608} &\textbf{0.0406} &\textbf{0.0914} &\textbf{0.0505}\\
  \hline
\end{tabular}
\end{table*}

\begin{figure}
    \centering
    \includegraphics[width=9cm]{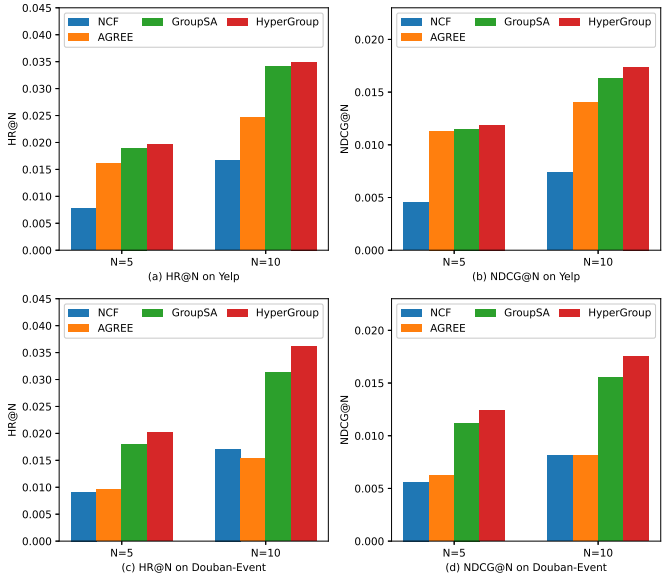}
    \caption{Top-N recommendation performance on individual users.}
    \label{fig:results_users}
\end{figure}

\section{Experimental Results (RQ1)}
The comparison results with the baseline methods are shown in Table~\ref{tab:results_all}, from which we can observe that: 1) \ac{HyperGroup} significantly outperforms all the baselines on the two datasets (all the improvements are statistically significant with $p<0.01$), which demonstrates the advantage of our hyperedge embedding-based solution.
2) NCF performs better than BPR-MF, but it can only get similar or even worse results than Pop. This is because in \ac{OGR}, groups are formed occasionally, and the observed group-item interactions are extremely sparse. In our Yelp and Douban-Event datasets, the average numbers of interactions per group are 1.12 and 1.47, respectively. This data sparsity issue makes it infeasible to treat a group as a virtual user and learn a group's interests only from her historical group-item interaction data (i.e., NCF and BPR-MF).
3) The performance of group recommendation methods (PIT, COM, AGREE, SIGR, GroupSA and \ac{HyperGroup}) developed for \ac{OGR} achieves superior performance over the recommendation methods proposed for individual users (i.e., NCF and Pop). This again demonstrates the complexity of the occasional group recommendation process, and simply view groups as virtual users cannot get satisfactory results. 4) Neural network-based group recommendation methods (i.e., AGREE, SIGR, GroupSA, GroupIM and \ac{HyperGroup}) outperform topic model-based methods (PIT and COM), indicating the capability of neural networks in capturing group members' behaviour patterns, which can lead to a more accurate recommendation result. COM outperforms PIT because groups in the two datasets are loosely organized and there may not exist a representative member to make item selections for a group. 5) The methods that consider user's social influence (\ac{HyperGroup}, GroupSA and SIGR) achieve better results than other baselines, demonstrating the benefits brought by exploiting the social influence.

To investigate the recommendation performance of our method in ranking items for individual users, we further compare \ac{HyperGroup} with baselines that can make recommendations for individuals (i.e., NCF, AGREE and GroupSA) on the user-item recommendation task.
% As GroupIM can not directly recommend items for users, we do not compare with it.
The experimental results are reported in Fig.~\ref{fig:results_users}, from which we can find that \ac{HyperGroup} also achieves the best performance, demonstrating the effectiveness and advantage of \ac{GNN} in learning the representations of individual users, as well as the joint training method to mutually enhance group recommendation and individual recommendation.

\begin{figure}[h]
    \centering
    \includegraphics[width=9cm]{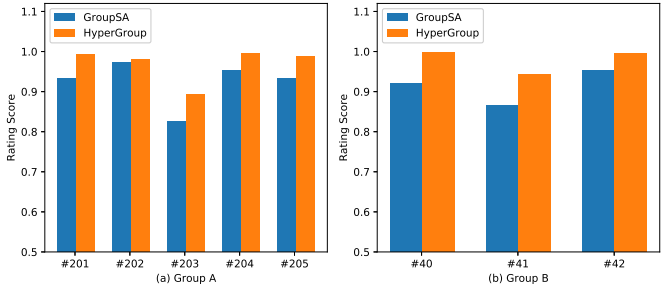}
    \caption{Case studies: preference scores predicted by \ac{HyperGroup} and GroupSA on Yelp.}
    \label{fig:case_study}
\end{figure}

\textbf{Case Studies.} Besides the above macro-level analysis, we also conduct case studies from a micro-level view via visualizing the rating scores for two randomly-chosen groups (A and B) from Yelp, which have interacted with items (\#201, \#202, \#203, \#204 and \#205) and items (\#40, \#41, \#42), respectively.  
To demonstrate the superiority of \ac{HyperGroup} in predicting the preferences of groups to items, we compare it with the state-of-the-art group recommendation method GroupSA. 
% As our method exploits an average strategy to aggregate the individual preferences learned in \ac{IPM}, we do not visualize the user weights here.
The experimental results are shown in Fig.~\ref{fig:case_study}, from which we find that \ac{HyperGroup} can make more accurate predictions than GroupSA, since for these ground-truth items, the predicted scores by \ac{HyperGroup} are more close to the target value $1$ than GroupSA. This result demonstrates the capability of \ac{HyperGroup} in predicting groups' preferences and thus leads to a better recommendation result for occasional groups.
\section{Model Analysis\label{model_analysis}}

In this section, we first conduct ablation studies to investigate the impact of model components and hyper-parameters to \ac{HyperGroup}. Then, the training efficiency of \ac{HyperGroup} is further investigated.

\begin{table*}
  \caption{Importance of components of \ac{HyperGroup}.}
  \label{tab:components}
  \centering
%   \scriptsize
% \footnotesize
  \begin{tabular}{|c|c|c|c|c|c|c|c|c|}
    \hline
    \multicolumn{9}{|c|}{Importance of Components of \ac{HyperGroup}} \\
    \hline
    \multirow{3}*{Methods} &\multicolumn{4}{c|}{Yelp}
    &\multicolumn{4}{c|}{Douban-Event}\\
    \cline{2-9}
    & \multicolumn{2}{c|}{$N$=5} &\multicolumn{2}{c|}{$N$=10}
    & \multicolumn{2}{c|}{$N$=5} &\multicolumn{2}{c|}{$N$=10}\\
    \cline{2-9}
    &HR & NDCG & HR& NDCG &HR & NDCG & HR& NDCG \\
    \hline
    AGREE &0.0569 & 0.0389 & 0.0896 &0.0495
    &0.0122& 0.0073 &0.0255 &0.0116\\
    \hline
    HGroup-SH &0.2159 &0.1651 &0.2806 &0.1860
    &{0.0309} &{0.0225} &{0.0478} &{0.0280}\\
    \hline
    HGroup-S &{0.2679} &{0.2089} &{0.3354} &{0.2305}
    &{0.0384} &{0.0286} &{0.0545} &{0.0337}\\
    \hline
    HGroup-H &0.3543 &0.2764&0.4339 &0.3021
    &0.0514 &0.0340  &0.0789&0.0429\\
    \hline
     \ac{HyperGroup} &\textbf{0.4827} &\textbf{0.3973} &\textbf{0.5598}&\textbf{0.4223}
     &\textbf{0.0608} &\textbf{0.0406} &\textbf{0.0914} &\textbf{0.0505}\\
    \hline
\end{tabular}
\end{table*}

\subsection{Importance of Components (RQ2)}

To investigate the importance of  \ac{IPM} and \ac{HRL}, we first compare \ac{HyperGroup} with its three variants:
\begin{itemize}
     \item \textbf{HGroup-SH.} This is a simplified version of \ac{HyperGroup} that replaces both the \ac{IPM} and \ac{HRL} components with more basic components. Specifically, it uses only a basic matrix factorization model without consideration of the social network information to learn individual preferences of group members, and then employs a vanilla attention-based preference aggregation strategy to learn groups' preferences. This is to study the effectiveness of these two components.
    \item \textbf{HGroup-S.} This is another variant of \ac{HyperGroup} that removes the \ac{IPM} component, that is, excluding the \ac{GNN}-based individual preference learning mechanism from \ac{HyperGroup}, and utilizes a basic matrix factorization model to learn users' personal preferences. This is to validate the importance of leveraging friends' preferences to enhance the individual's preference learning.
    \item \textbf{HGroup-H.} This variant removes the \ac{HRL} component from \ac{HyperGroup}, and utilizes an average aggregation-based strategy to learn the groups' representations by averaging the personal preferences of group members. This is to evaluate the effect of our hyperedge embedding-based group preference learning mechanism.
\end{itemize}

\begin{table*}
  \caption{Importance of User-item Interaction Data.}
  \label{tab:joint}
  \centering
%   \scriptsize
% \footnotesize
  \begin{tabular}{|c|c|c|c|c|c|c|c|c|}
    \hline
    \multicolumn{9}{|c|}{Importance of User-Item Interaction Data} \\
    \hline
    \multirow{3}*{Methods} &\multicolumn{4}{c|}{Yelp} 
    &\multicolumn{4}{c|}{Douban-Event}\\
    \cline{2-9}
    & \multicolumn{2}{c|}{$N$=5} &\multicolumn{2}{c|}{$N$=10}& \multicolumn{2}{c|}{$N$=5} &\multicolumn{2}{c|}{$N$=10}\\
    \cline{2-9}
    &HR & NDCG & HR& NDCG &HR & NDCG & HR& NDCG\\
    \hline
    NCF &0.0110 &0.0074 &0.0193 &0.0100
    &0.0041 &0.0024 &0.0061 &0.0030\\
    \hline
    HGroup-U &0.0123 &0.0081 &0.0201 &0.0106
    &0.0031 &0.0019 &0.0054&0.0026\\
    \hline
    \ac{HyperGroup} &\textbf{0.4827} &\textbf{0.3973} &\textbf{0.5598}&\textbf{0.4223}
    &\textbf{0.0608} &\textbf{0.0406} &\textbf{0.0914} &\textbf{0.0505}\\
  \hline
\end{tabular}
\end{table*}

\subsubsection{Importance of \ac{IPM}}
To evaluate our individual preference learning component, we compare \ac{HyperGroup} with HGroup-SH and HGroup-S. The experimental results are reported in Table~\ref{tab:components}, from which we have the following observations: (1) \ac{HyperGroup} significantly outperforms HGroup-SH and HGroup-S on both datasets, indicating the importance of the \ac{IPM} and \ac{HRL} components, and only considering one of them alone cannot get better results than combing them together. (2) \ac{HyperGroup} performs better than HGroup-S, demonstrating the benefit of exploiting the social interests to alleviate the data sparsity of user-item interactions and the effectiveness of our
\ac{GNN}-based learning paradigm. This result also demonstrates that \ac{IPM} is able to provide solid foundations for further learning group representations via hyperedge embedding techniques. (3) From Table~\ref{tab:components}, we also notice that all our variants of \ac{HyperGroup} perform better than AGREE which only utilizes the attention mechanism to make group recommendation. This result demonstrates the effectiveness of our \ac{HyperGroup} solution and the importance of exploiting social connections to alleviate the sparsity issue of users' interaction data as well as the benefit of investigating group similarity to enhance groups' preference learning.

\subsubsection{Importance of \ac{HRL}} To validate our \ac{HRL} component, we further conduct another ablation study by comparing \ac{HyperGroup} with HGroup-H. From the experimental results shown in Table~\ref{tab:components}, we can observe that: (1) \ac{HyperGroup} outperforms HGroup-H demonstrating the power of our hierarchical neural network and the effectiveness of learning group representations by treating groups as hyperedges in a hypergraph. That is, modeling the group similarity in terms of common group members is helpful for learning a better group representation and thus leads to a better group recommendation. (2) We also notice that there is a bigger gap between \ac{HyperGroup} and HGroup-S than that between \ac{HyperGroup} and HGroup-H, which indicates that without effective individual preferences, we can only get sub-optimal group representations, and accurate personal preferences of group members can provide foundations for learning effective group representations.

\subsubsection{Importance of the User-item Interactions}
To validate the importance of integrating user-item interaction data to enhance the group preference learning process, we conduct experiments to compare \ac{HyperGroup} with NCF and HGroup-U:
\begin{itemize}
    \item \textbf{HGroup-U.} This is a variant of \ac{HyperGroup} that does not integrate the user-item interaction data and only utilizes the group-item interaction data to train the model.
\end{itemize}

The experimental results are reported in Table~\ref{tab:joint}. From the results we have the following observations: (1) \ac{HyperGroup} consistently and significantly outperforms HGroup-U in both datasets, which validates the usefulness of user-item interaction data in enhancing the training of our group recommendation model, that is, leveraging the learned user and item representations from user-item interactions to provide solid foundations for group preference learning.
(2) HGroup-U outperforms NCF, which demonstrates the capability of our method in learning group preferences, that is, the advantage of the \ac{GNN}-based hyperedge embedding method. This result also demonstrates the importance of exploiting the group similarity in group representation learning. But we also notice that without the help of user-item interactions, the performance of HGroup-U has a big gap with \ac{HyperGroup}, due to the extreme sparsity of group-item interaction data.

\begin{figure}
    \centering
    \includegraphics[width=9cm]{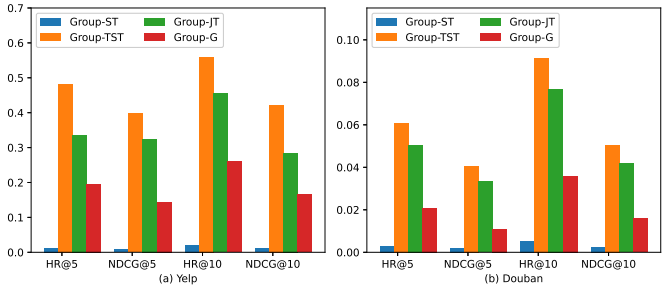}
    \caption{Comparison of different model optimization approaches.}
    \label{fig:optimization}
\end{figure}

\subsubsection{Comparison of Different Model Optimization Approaches}
To evaluate the performance of our proposed model optimization strategies on heterogeneous data, we compare the following four model optimization strategies: 
\begin{itemize}
    \item \textbf{Group-ST.} This is the single training strategy that optimizes \ac{HyperGroup} only on the group-item interactions, which is also known as HGroup-U.
    \item \textbf{Group-G.} This is the strategy that optimizes \ac{HyperGroup} via only considering the user-item loss.
    \item \textbf{Group-TST.} This strategy integrates the user-item interactions and group-item interactions via a two-stage training method (as shown in Section~\ref{model_optimization}).
    \item \textbf{Group-JT.} This is another optimization method that integrates the user-item interaction data via jointly training the group-item and user-item recommendation task simultaneously (as shown in Section~\ref{model_optimization}).
\end{itemize}

From the experimental results reported in Fig.~\ref{fig:optimization} we can observe that: (1) All joint training methods (i.e., Group-TST and Group-JT) significantly outperform the single training methods Group-ST and Group-G, demonstrating the significance of leveraging the user-item interaction data for training \ac{HyperGroup}, and the ability of our joint model optimization method in addressing the heterogeneous data (i.e., the mixture of user-item and group-item interaction data). (2) Group-TST performs better than Group-JT, demonstrating the two-stage training method is more suitable to optimize our hierarchical group recommendation model, which first learns the embeddings of individual users in the lower layer of \ac{HyperGroup} and then based on that learns group embeddings in the higher layer of our model.

\begin{table}
    \centering
    \caption{Impact of parameter $S$.
    }
    % \footnotesize
    \begin{tabular}{c|c|c|c|c}
    \hline
         $S$ &HR@5 &NDCG@5 &HR@10 &NDCG@10\\
    \hline
        1 &0.4339 &0.3531 &0.5160 &0.3796 \\
    \hline
        2&0.4459&0.3612 &0.5203 &0.3853 \\
    \hline
        3 &0.4452 &0.3670 &0.5314&0.3950\\
    \hline
        4&0.4827 &0.3973 &0.5598 &0.4223\\
    \hline
        5&0.4734 &0.3873 &0.5517 &0.4126\\
    \hline
    \end{tabular}
    \label{tab:parameter_S}
\end{table}
% \vspace{-0.5cm}
\begin{table}
    \centering
    \caption{Impact of parameter $N_x$.}
    % \footnotesize
    \begin{tabular}{c|c|c|c|c}
    \hline
         $N_x$ &HR@5 &NDCG@5 &HR@10 &NDCG@10\\
    \hline
            1 &0.4827 &0.3973 &0.5598 &0.4223\\
    \hline
        2&0.5582&0.4797 &0.6254 &0.5017 \\
    \hline
        3&0.5862&0.4963 &0.6600 &0.5201\\
    \hline
        4&0.5845&0.5030 &0.6440 &0.5223\\
    \hline
        5& 0.5842&0.4866 &0.6594 &0.5110\\
    \hline
    \end{tabular}
    \label{tab:parameter_Nx}
\end{table}
% \vspace{-0.5cm}
\begin{table}
    \centering
    \caption{Impact of parameter $w$.}
    % \footnotesize
    \begin{tabular}{c|c|c|c|c}
    \hline
         $w$ &HR@5 &NDCG@5 &HR@10 &NDCG@10\\
    \hline
        0.1 &0.4452 &0.3670 &0.5314&0.3950\\
    \hline
        0.3&0.4736&0.3870 &0.5505 &0.4119 \\
    \hline
        0.5&0.4827 &0.3973 &0.5598 &0.4223\\
    \hline
        0.7&0.4554&0.3754 &0.5307 &0.3998\\
    \hline
        0.9& 0.4541&0.3682 &0.5323 &0.3936\\
    \hline
    \end{tabular}
    \label{tab:parameter_w}
\end{table}

\begin{table}
    \centering
    \caption{Impact of Different Group Sizes (denoted by $l$ ).}
    % \footnotesize
    \begin{tabular}{c|c|c|c|c}
    \hline
         $l$ &HR@5 &NDCG@5 &HR@10   &NDCG@10\\
    \hline
        $l<3$& 0.4709 &0.3779&0.5467 &0.4023 \\
    \hline
        $3\leq l \leq 7$&0.5529 &0.4848&0.6020 &0.5007 \\
    \hline
        $7 < l$&0.4580 &0.3770&0.5526 &0.4073\\
    \hline
    \end{tabular}
    \label{tab:group_size}
\end{table}

\begin{table}
    \centering
    \caption{Top-$N$ Recommendation Performance on Yelp ($\tau$ denotes the number of items visited by groups).}
    % \footnotesize
    \begin{tabular}{c|c|c|c|c}
    \hline
         $\tau$ &HR@5 &NDCG@5  &HR@10 &NDCG@10\\
      \hline
        $ \tau \leq 3$ &0.4668 &0.3633&0.5576 &0.3927\\
    $\tau >3 $ &0.4936 &0.4164 &0.5684 &0.4408 \\ 
    \hline
    \end{tabular}
    \label{tab:group_rating_number}
\end{table}

\subsection{Impact of Hyper-parameters (RQ3)\label{hyperparameters}}

Tables~\ref{tab:parameter_S}-\ref{tab:parameter_w} present the experimental results on tuning the hyper-parameters of \ac{HyperGroup}. Due to similar results are achieved on Douban-event, only the results on Yelp are reported.

\subsubsection{Impact of $S$}
The hyper-parameter $S$ refers to the number of neighbors sampled at each layer. 
A higher value of $S$ indicates that there are more neighbors of users or groups that are aggregated in the corresponding aggregation functions. 
The recommendation performance with respect to $S$ is shown in Table~\ref{tab:parameter_S} (same values of $S$ for both components are utilized in this result), from which we find diminishing returns for sampling large neighbors, and when the number of sampled neighbors surpass a certain value, the recommendation performance will even deteriorate, since more unrelated users or groups are considered. Moreover, we also notice that large sampled neighbors significantly increase the running time. To strike a balance between running time and performance, we set $S=4$ for both \ac{IPM} and \ac{HRL} components of \ac{HyperGroup} on two datasets. 

\subsubsection{Impact of $N_x$}
We investigate the performance of \ac{HyperGroup} with respect to different values of $N_x$, which denotes the number of negative samples utilized for per positive sample. As the results shown in Table~\ref{tab:parameter_Nx}, there is a high variance induced by the number of sampled negative examples. From this result, we can find that generating more negative samples for per positive sample is helpful to obtain a more accurate recommendation model. The best performance of our method is achieved when $N_x=4$. This result also indicates that very few negative samples can already lead to satisfactory results.
In experiments, we set $N_x=1$ on both Yelp and Douban-Event as in~\cite{guo2020group} to make our results comparable with them.

\subsubsection{Impact of $w$}
To explore the impact of the residual connection (i.e., the hyper-parameter $w$ in Eq.(\ref{eq:residual})), we further conduct experiments by varying the values of $w$, which plays a role control the contributions of the two types of group representations. The experimental results are shown in Table~\ref{tab:parameter_w}. The best performance is achieved at $w=0.5$ on Yelp, and $w=0.3$ on Douban-Event, from which we can observe that if we pay less attention to the individual preferences, we cannot get a better group representation, since the group preference learning takes the preferences of individual group members as foundations.

\subsubsection{Impact of $l$} 
To study the performance of our method on groups with different sizes, we evaluate \ac{HyperGroup} by splitting groups in the testing data into three bins based on their size, that is, small ($l<3$), medium ($3\le l \le 7$) and large ($l>7$). The experimental results are reported in Table~\ref{tab:group_size}, which indicate that \ac{HyperGroup} is more suitable to make recommendations for medium groups. The best result is achieved on the medium group bin (i.e., $3\le l \le 7$). The main reason behind this result is that in a
small group we do not have enough group-user interactions to explore the group-level similarities, and in a large group, the group members are more difficult to reach consensus due to the personal interests of individuals.

\subsubsection{Impact of $\tau$}
To test our model's performance on different levels of item interaction sparsity (cold-start vs. popular items), we conduct experiments on items with different activity levels on Yelp, where group-item interactions in the test data are split into two bins based on item activity, that is, interactions with cold-start items ($\tau \leq 3$) and interactions with popular items ($\tau >3$). The experimental results are reported in Table~\ref{tab:group_rating_number}, from which we can observe that \ac{HyperGroup} achieves expected performance on popular items. But we also find that \ac{HyperGroup} achieves comparable results on cold-start items, which demonstrates the importance of group members' individual preferences and social interests in recommending items to occasional groups, as well as the effectiveness of our hierarchical hyperedge embedding-based solution.

\begin{figure}
    \centering
    \includegraphics[width=12cm]{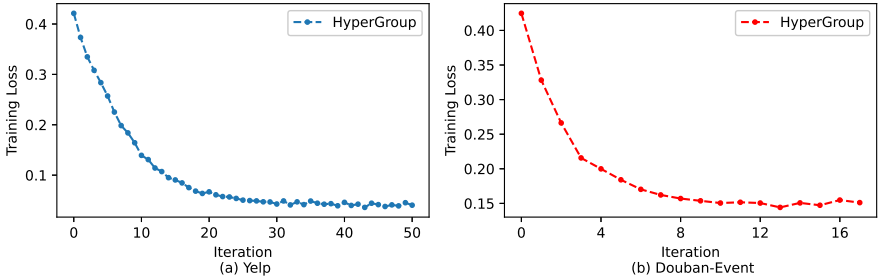}
    \caption{Training loss of HyperGroup w.r.t the number of iterations on Yelp and Douban-Event.}
    \label{fig:training_loss}
\end{figure}

\subsubsection{Convergence} 
To demonstrate the rationality of our learning scheme, we report the value of training loss along with each iteration using the optimal parameter setting in Fig.~\ref{fig:training_loss}. From this result, we can observe that with the increasing number of iterations the training loss of HyperGroup gradually decreases on both datasets. On Yelp HyperGroup converges fast in the first 20 iterations, and reaches its optimal results around the 30th iteration, while on Douban-Event it achieves its best performance around the 10th iteration. This result indicates the rationality of our training strategy.

We also explore the impact of different feature generation methods utilized in the \ac{IPM} component (i.e., the features denoted by $\boldsymbol{x}_u$), but the results achieved by different initialization strategies for embedding features are very close after the \ac{HyperGroup} model is fully trained. Hence, we do not report these results.

\begin{figure}
    \centering
    \includegraphics[width=12cm]{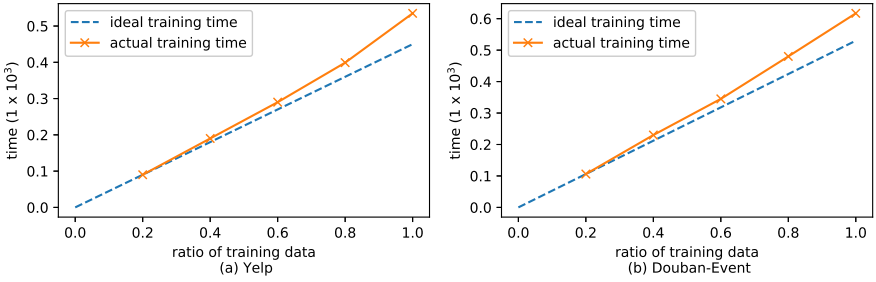}
    \caption{Training time of \ac{HyperGroup} with different data sizes.}
    \label{fig:training_time}
\end{figure}

\subsection{Training Efficiency and Scalability (RQ4)}

To investigate the practicality of our recommendation method in real-world applications, we validate the training efficiency and scalability of \ac{HyperGroup} via measuring the time cost for the model training with different proportions of the training data (Yelp and Douban-Event). That is, we vary the ratios of the training data in $\{0.2, 0.4, 0.6, 0.8, 1.0\}$, and then report the corresponding training time in Fig.~\ref{fig:training_time}. The experimental results of \ac{HyperGroup} are obtained with all the hyper-parameters
are fixed.
To make our results comparable, the expected ideal training time that is linearly associated with the number of training samples is also reported in Fig.~\ref{fig:training_time}.
From the experimental results, we can observe that when the ratio of the training data gradually increases from 0.2 to 1.0, the time cost for training \ac{HyperGroup} on Yelp grows from $0.09\times 10^3$ seconds to $0.535 \times 10^3$ seconds and it grows from $0.106 \times 10^3$ seconds to $0.617\times 10^3$ seconds on Douban-Event. The overall trend on these two datasets shows that the dependency of times cost for training \ac{HyperGroup} on the data scale is approximately linear. 
This result provides us positive evidence to answer RQ4, that is, \ac{HyperGroup} is scalable
to large scale datasets.
\section{Conclusions}
In this work, we investigated the \ac{OGR} problem, and proposed a hierarchical \ac{GNN}-based group recommender \ac{HyperGroup} to learn the group preference via the hyperedge embedding technique based on the learned individual preferences of group members.
In this way, our method not only can model the individual-level preferences, but also the group-level communications.
Specifically, to alleviate the sparsity issue of user-item interactions, we first learned group members' personal preferences by leveraging their social interests to provide solid foundations for group representation learning.
Then, to enhance the group representations by leveraging the group similarity, we connected all groups as a hypergraph, and proposed a hyperedge embedding method to solve the \ac{OGR} problem in the higher-layer of our network.
Finally, to leverage the user-item interactions to further accelerate the training process of the group-item recommendation task, two joint optimization strategies were developed.
To validate the effectiveness of our \ac{HyperGroup}, we conducted extensive experiments on two real-world datasets that are proposed for \ac{OGR} task. The experimental results demonstrated the superiority of our hierarchical hyperedge embedding-based solution in making recommendations for occasional groups.

Besides the Yelp and Douban-Event datasets, our method can also be applied to other real-world settings, such as the users who attend an academic conference or the friends that meet at social events. As in the above cases where groups are formed occasionally, there are no historical group activities. Recommending items (such as trips or restaurants) to these kinds of groups falls into the \ac{OGR} scenario, where our \ac{HyperGroup} method can be applied. That is, first learning the interests of individual users by leveraging their social connections, and then inferring groups' representations via the overlapping relationship among them.


\begin{acks}
This work was supported by National Natural Science Foundation of China (No. 61602282), ARC Discovery Project (No. DP190101985) and China Postdoctoral Science Foundation (No. 2016M602181).
\end{acks}

\bibliographystyle{ACM-Reference-Format}
\bibliography{references}

\end{document}